\documentclass[aps,prl,twocolumn,groupedaddress]{revtex4}

\usepackage{natbib}
\usepackage{graphicx}
\usepackage{amsmath,amssymb}
\usepackage{ulem}

\newcommand{\be}{\begin{equation}}
\newcommand{\ee}{\end{equation}}

\begin{document}

\title{Quasi-isotropic cascade in MHD turbulence with mean field}
\author{Roland~Grappin}
\email[]{Roland.Grappin@obspm.fr}
\affiliation{LUTH, Observatoire de Paris and LPP, Ecole Polytechnique
}

\author{Wolf-Christian~M\"{u}ller}
\email[]{Wolf.Mueller@ipp.mpg.de}
\affiliation{Max-Planck Institut f\"ur Plasmaphysik, 85748 Garching, Germany}

\author{\"Ozg\"ur G\"urcan}

\affiliation{LPP, Ecole Polytechnique}

\date{\today}

\begin{abstract}
We propose a phenomenological theory of incompressible magnetohydrodynamic turbulence in the presence of a strong large-scale magnetic field,
which establishes a link between the known anisotropic models of strong and weak MHD turbulence
We argue that the Iroshnikov-Kraichnan isotropic cascade develops naturally within the plane perpendicular to the mean field, while
oblique-parallel cascades with weaker amplitudes can develop, triggered by the perpendicular cascade, with a reduced flux resulting from a quasi-resonance condition.
The resulting energy spectrum $E(k_\parallel,k_\bot)$ has the same slope in all directions. 
The ratio between the extents of the inertial range in the parallel and perpendicular directions
is equal to $b_{rms}/B_0$.
These properties match those found in recent 3D MHD simulations with isotropic forcing reported in [R. Grappin and W.-C. M\"uller, Phys. Rev. E $\bold{82}$, 26406 (2010)].
\end{abstract}
\maketitle

\textit{Introduction} - 
Plasma turbulence in a mean magnetic field corresponds to a rather ubiquitous astrophysical setting. 
Understanding its nonlinear dynamics which gives rise to measurable two-point statistics such as the energy spectrum is thus highly important. 
The current state of affairs with regard to the simplified incompressible magnetohydrodynamic approximation may be summarized as follows.
MHD turbulence with strong mean field $B_0$ mainly forms gradients in directions perpendicular to the mean field,
as the nonlinear cascade is based on interactions of Alfv\'en waves propagating in opposite directions along the mean field.
These interactions are resonant, and thus most efficient, perpendicularly to the mean field
\cite{Montgomery:1981p6103,Shebalin:1983p6056,Grappin:1986p707}.
Predicting the exact form of anisotropy and the corresponding scaling of 3D spectra is no simple problem.

The first phenomenology proposed for MHD turbulence was a weak version of Kolmogorov phenomenology,
based on non-resonant coupling between oppositely propagating Alfv\'en waves, and ignoring anisotropy
 \cite{Iroshnikov:1963p9274,Kraichnan:1965p9279} (IK). 
It predicted a reduced energy flux and an isotropic spectrum with a spectral slope $-3/2$ instead of the $-5/3$ Kolomogorov slope.
It has been shown since to hold in 2D MHD with the rms magnetic field playing the role of the mean field \cite{Pouquet:1988dv,Biskamp:1989ug}.

Later on, it was proposed that 3D MHD turbulence with guide field would develop along 
either purely perpendicular directions (weak turbulence, \cite{2000JPlPh..63..447G,Verdini:2012hu})
or quasi-perpendicular directions (strong turbulence, \cite{Goldreich:1995p4882,Boldyrev:2006p4917}),
depending on the importance of the Alfv\'en wave period $t_a = 1/k_\parallel B_0$ compared to the perpendicular nonlinear turnover time $t_{NL}=1/kb \simeq 1/k_\bot b$.
In the strong case, anisotropy is ruled by the so-called critical balance (CB) between both times $\chi = t_a/t_{NL} = k_\bot b/k_\parallel B_0=1$, 
which corresponds to quasi-resonant couplings and limits
the growth of parallel wave numbers during the cascade in wave number space, while in the weak turbulence case, one has $\chi \ll 1$.

The work presented here is meant to solve two issues. First, the two theories of weak turbulence, the isotropic one (that applies to 2D, \cite{Iroshnikov:1963p9274,Kraichnan:1965p9279}) and the anisotropic one (3D, \cite{2000JPlPh..63..447G})
are not compatible, as the first one has a spectral slope $-3/2$ and the second one a slope $-2$.
It is however expected that one should pass from the 3D to 2D when sufficiently increasing the mean field amplitude $B_0$ (see also  \cite{Muller:2003p809}).
Second, a recent DNS study has reported a 3D regime with mild anisotropy and uniform $-3/2$ spectral slope in all directions \citep{2010PhRvE..82b6406G} (GM).
In this regime, the extent of the inertial range remains independent of $B_0$ in the perpendicular directions, and varies as $1/B_0$ in the parallel direction.
The purpose of this paper is thus to build a phenomenology of 3D MHD turbulence which tends to the IK regime in the limit $B_0 \rightarrow \infty$, 
thus sharing the properties of the GM regime.

Some authors \cite{Beresnyak:2012ek,Chen:2011cv} argued that the GM results are an artefact due to measuring the spectrum in a global frame,
while the true anisotropy can be obtained only by measuring spectra in a frame attached to the local magnetic field.
While agreeing that the local frame of reference is the physical spectral symmetry axis, we think that still in the GM regime
the properties are identical in the local and global frame, basically because the effect of frame change amounts to shuffling the symmetry axis by an angle of about
$b_{rms}/B_0 = 1/5$ which is too small to change a quasi-perpendicular spectrum into the GM spectrum, and cannot substantially modify the GM spectrum itself which fills the whole angular range.

We propose in this Letter a model of the GM regime combining an IK cascade in the perpendicular plane, coupled with a slave cascade in the oblique and parallel directions. 
In the perpendicular plane, the IK cascade is based on the $b_{rms}$ field, which plays the role of the local mean field. 
Critical balance is assumed to develop during a first phase, ensuring a minimum extent of the spectrum in the parallel direction. 
Then, starting with the forced parallel and oblique modes not satisfying CB (i.e. with relatively large $k_\parallel$), 
a slave cascade can begin. 
The model describes the evolution of fluctuations which do not satisfy the standard CB criterion; it
consists in a flip-flop or ricochet process between the parallel and oblique wave vectors directions, 
each ricochet corresponding to an increase in either the oblique or parallel wave vectors (see below Fig.~\ref{fig3}). 
A quasi-resonance condition,
which can be regarded as a generalization of the CB criterion, 
limits the increase at each ricochet.
This leads to a reduced, IK-like energy flux.
An isotropic scaling results, and as a corollary, the ratio between the parallel and perpendicular inertial range extents is $b_{rms}/B_0$, as obtained numerically in GM.

\textit{Equations and basic models - } We start with the isotropic Euler case. Formally, the Euler equation can be written in Fourier space as
\be
\partial_t u_k = \Sigma_{p,q} k u_p u_q
\label{Euler}
\ee
It is implicit that in the r.h.s. a sum is made on wave vectors $p$ and $q$ such that $\mathbf{k=p+q}$. 
The equation omits to mention vector components and a complicated kernel has been replaced by the (dimensionally correct) factor $k$.
If only a single triad $(k, p, q)$ is retained in the r.h.s. of eq.~\ref{Euler}, no spectrum will form, but taking two nearby triads opens the possibility of a cascade. 
This is the basis of the so-called shell models starting with the Desnyansky-Novikov equations  \cite{Desnyansky1974}
in which one uses a discretization of wavenumber space into shells with mean radius growing in a geometric progression: 
$k_n = 2^n k_0$. 
A still more simplified version of this model amounts to write down a flux equation for the energy in each shell of wave numbers,
with $E_k$ being the 1D energy spectrum:
\begin{equation}
u_n^2/2  \simeq kE_k
\end{equation}
The variable $u_n = u(k_n)$ represents all the modes within shell number $n$. The system reads
\be
\partial_t u_n^2 = F_n - F_{n+1}
\label{DNbis}
\ee
where the energy flux in wave number space is
\be
F_n = k_n u_n^3
\ee
This system admits a stationary solution with the Kolmogorov scaling $u \propto k^{-1/3}$
and the characteristic cascade time $t_{tr}(k)$ of the cascade from $k_n$ to $k_{n+1}$ is
\begin{equation}
t_{tr}^{-1}=F/u^2 = ku
\label{tNL}
\end{equation}

We now consider the MHD case, in which wave propagation decreases the coupling along parallel wave numbers,
compared to the coupling in the previous isotropic case.
The MHD equations in Fourier space read formally, written in terms of
the so-called Elsasser variables $u^\pm = u\mp b$:
\begin{equation}
\partial_t u^\pm_k \pm i k.B_0 u^\pm_k = \Sigma_{k=p+q} k u^\pm_p u^\mp_q
\label{base1}
\end{equation}
This is the analoguous to eq.~\ref{Euler}.
Changing variables
\begin{equation}
u^\pm_k = U^\pm_k e^{\pm i k.B_0 t}
\end{equation}
allows to rewrite eq.~\ref{base1} in terms of the wave amplitudes $U^\pm$ (using $k=p+q$) as
\begin{equation}
\partial_t U^\pm_k  = \Sigma_{k=p+q} \ k \ U^\pm_p U^\mp_q e^{\mp 2i q.B_0 t}
\label{base2}
\end{equation}
(See  \cite{Grappin:1986p707}) for an exact form of the equations in the 2D case).
In the following we do not distinguish further $U^+$ from $U^-$ and use $u$ for both fields.

\textit{Quasi-resonance - }
We start initially with a spectrum concentrated at large scales.
To examine which triads allow to progress along the $Ox$ mean field axis, we
assume that only quasi-resonant couplings are non-negligible,
more precisely that:
\begin{equation}
q_x B_0 \le T_{cor}^{-1}
\label{BC1}
\end{equation}
where $T_{cor}$ is the correlation time of $u_pu_q$, that is $T_{cor}^{-1} = \max(pu_p, qu_q$).
Assuming that nonlinear coupling are local:
\begin{eqnarray}
p \simeq q \simeq k \label{tr4}
\end{eqnarray}
and also (which is always true) that the perpendicular spectrum is dominant, we find:
\begin{equation}
T_{cor} Ê\simeq 1/(qu_q)
\label{Tcor}
\end{equation}
The quasi-resonance condition (eq.~\ref{BC1}) thus reads
\begin{equation}
q_x B_0 \le qu_q
\label{BC2}
\end{equation}
In the following we adopt:
\begin{equation}
q_x \simeq qu_q/B_0
\label{BC3}
\end{equation}
Standard CB phenomenology can be viewed as an application of eq.~\ref{BC3} to all vectors of each triad: $k_x \simeq ku_k/B_0$. 
A representative triad lies within a plane inclined with respect to the plane $yOz$ by a small angle
$\simeq u_k/B_0$. To be able to escape this quasi-perpendicular cascade, we need now (i) an initial seed well outside the quasi-resonance ``cone'' $q_x/q = u_q/B_0$ (ii) a way to build growing triads in the ($k_\|, k_\bot$) plane.


\textit{The oblique and parallel cascades}
We now reason in the same way as when dealing with eq.~\ref{DNbis} above.
There we considered an isotropic cascade, assuming a progression towards larger wave number in discrete steps from $k_n$ to $k_{n+1}$.
Here we look for a cascade along the $Ox$ axis but with a slower and slower pace, 
due to the quasi-resonance constraint. How much slower?

To answer, we first consider only perfectly resonant triads, hence with $q_x=0$ (orthogonal triad).
Such triads do not lead to any progression along $Ox$.
Indeed, the triad assumes that the mode $u_p$ already exists, with a wave vector that has already an $x-$component equal to that of the final wave vector.

\begin{figure}
\includegraphics [width=7cm]{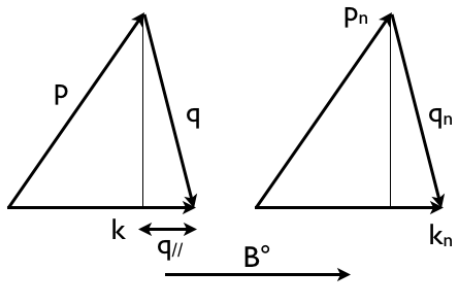}
\caption{
The basic triad $\mathbf{k=p+q}$ for the parallel (k) and oblique (p) cascades, with the quasi-perpendicular wave vector (q).
}
\label{fig1}
\end{figure}

\begin{figure}
\includegraphics [width=8cm]{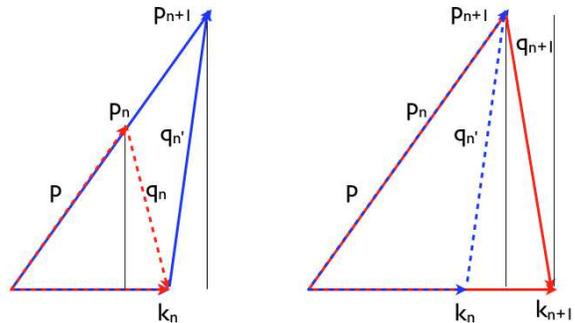}
\caption{
Sketches of the two basic steps of the oblique cascade (left) and parallel cascade (right) from the large scales into the unexcited part of the spectral space.
Left: From $p_n$ to $p_{n+1}$ (a) triad in dashed red transfers excitation from $p_n$ to $k_n$.
(b) the (blue solid) triad transfers excitation from $k_n$ to $p_{n+1}$.
Right: From $k_n$ to $k_{n+1}$ 
(a) triad in dashed blue transfers excitation from $k_n$ to $p_{n+1}$.
(b) triad in solid red transfers excitation from $p_{n+1}$ to $k_{n+1}$.
}
\label{fig2}
\end{figure}

\begin{figure}
\includegraphics [width=6cm]{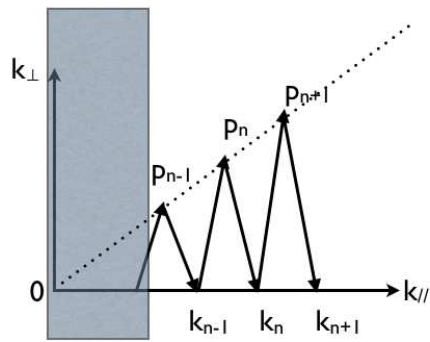}
\caption{
Ricochet process: successive quasi-perpendicular wave vectors $q_n$ allowing the coupled cascade between the parallel and oblique directions,
each step being illustrated by a triad illustrated in the previous figure.
The quasi-perpendicular wave vectors $q_n$ and $q_n'$ satisfy the quasi-resonant condition compatible with critical balance.
The quasi-perpendicular spectrum is thus supposed to be available to drive the coupled parallel-oblique cascades.
The shaded area represents the initial spectral extent.
}
\label{fig3}
\end{figure}

Next we consider quasi-resonant triads as in eq.~\ref{BC3}, i.e. with $p_x \simeq k-q_x$ and
$q_x >0$.
Such triads allow to progress from 
$k_{n-1}=k-q_x$ up to $k_n = k$.
This is illustrated in figs.~\ref{fig1}-\ref{fig3}.
Figs.~\ref{fig1}-\ref{fig2} show that the same triad intervenes to drive the parallel and the oblique
cascades, while fig.~\ref{fig3} gives a global view of the cascade by showing how the quasi-perpendicular modes allow to progressively populate the plane along 
the parallel and the oblique axis.

In fig.~\ref{fig3}, the angle of the oblique path has been chosen to be about $\pi/4$. The important point is that the possibility to propagate the excitation using the triads shown
in the figure relies first on the fact that the oblique and quasi-perpendicular wave-vectors (reps. $p$ and $q$) are initially excited.
Then, as the quasi-perpendicular vector $q$ grows, it will necessarily come out of the initial rectangular gray region.
In fact, the continuation of the oblique cascade requires that, as $q$ grows, 
the parallel component $q_x$ grows as well.
Its maximum value is given by eq.~\ref{BC3}.
For such values of $q_x$ to be available, we need that the perpendicular spectrum develops (or has developed) according to the critical balance, filling a cone around the
perpendicular direction, with parallel width varying as eq.~\ref{BC3}.


\textit{Reduced flux in the oblique-parallel cascades - } 
Due to the quasi-resonance constraint, only triads with $q_x/q \le u_q/B_0$ contribute to the cascade along the parallel and oblique directions.
The main set of triads contributing to the cascade in the absence of mean field are characterized by $0 < q_x \lesssim q$.
Hence we see that the contributing subset of quasi-resonant triads is a fraction of order $q_x/q$ of the standard fraction.
We thus conclude that the energy flux in the oblique/parallel directions is to be reduced, compared to the unrestricted isotropic flux with no $B_0$, by a factor R:
\be
R \simeq q_x/q \simeq u_q/B_0
\ee
Starting from the expression of the energy flux with zero mean field written as $F = k u_k^2 u_q$ (see eq. 4), 
we take into account the reduction factor $R$ to obtain:
\be
F = k_n u_k^2 u^2_q/B_0
\label{FLFL}
\ee
The expression above for the flux implies that the transfer
time $t_{tr}^\parallel(k)$ for the parallel-oblique spectrum depends on the perpendicular spectrum as $t^\parallel_{tr}(k)^{-1} = F/u^2_k$ or
\begin{equation}
t^{\parallel+}_{tr}(k)^{-1}  = k u^-_q \times (u^-_q/B_0)
\label{tstar}
\end{equation}
A stationary cascade implies that the flux expression in eq.~\ref{FLFL} be constant. 
A consequence of eq.~\ref{FLFL} is that the sum of the two scaling indices for $u_k$ and $u_q$ is $-1/2$, or in other words the sum of the 1D slopes 
of the perpendicular and parallel-oblique spectra is $-3$.
(See \cite{1983A&A...126...51G,Pouquet:1988dv} for the case of $z^+$ and $z^-$ spectral slopes in isotropic IK turbulence).
This implies that the scaling will be isotropic if and only if the perpendicular slope is $3/2$.

If we remember the scenario developed when commenting Fig.~\ref{fig3}, 
the other basic requirement for the oblique/parallel cascades to proceed is that the CB is at work, i.e. that the quasi-perpendicular wave numbers 
do have a non-zero parallel component excited satisfying the CB. 
So we have shown that if the CB perpendicular cascade with 3/2 spectral scaling is at work, then oblique and parallel cascades with the same scaling are possible.

The existence of cascades restricted to the perpendicular direction only with $3/2$ spectral scaling is possible according to either one of two scenarios,
the first one being the IK scenario (as in 2D simulations) where non resonant-coupling is at work, 
the second one being restricted to quasi-resonant coupling but with local scaling of the velocity-magnetic field alignment weakening the interactions \citep{Boldyrev:2006p4917, Mason:2008p1}.
Re-examining our simulation data we found that the local alignment scaling is insufficient to explain the $3/2$ slope, so it is more likely that the IK scenario is at work in the GM case, in the 2D perpendicular plane.

The spectral extent, which is controlled by viscosity, provides a test of our phenomenology of coupled perpendicular and oblique cascades. 
Indeed, at the dissipative scale $1/k_d$, one should find equal transfer and dissipative times. 
This leads to two equalities close to one another, depending on whether we consider the parallel or perpendicular cascade:
\begin{eqnarray}
 \nu k^2  \simeq q u^2_q /B_0 Å \simeq k u^2_k / B_0
 \\
 \nu k^2  \simeq q u^2_q /b_{rms} Å \simeq k u^2_k / b_{rms}
\end{eqnarray}
hence
\be
k_{d_\parallel}/k_{d_\bot} = b_{rms}/B_0
\ee
This relation has been found numerically in GM.

We have proposed here a model of the GM regime in which the generation of the spectrum in the oblique and parallel directions is a genuine effect of the nonlinear
coupling between the oblique and parallel directions. The cascade is enslaved to the perpendicular cascade, but nevertheless the nonlinear coupling is there.
This scenario might not work for Reduced MHD (RMHD) which is often proposed as an equivalent but simpler
model for MHD with strong guide field. Indeed, in RMHD, any increase of the spectrum in the parallel direction is constrained by the CB condition $k_\parallel \le k_\bot b/B_0$.
The CB condition simply reflects the fact that the parallel gradients don't develop by their own nonlinear interactions but instead are formed by plain linear spatial transport
of the turbulent signals due to perpendicular eddies. This is not so in full MHD where parallel coupling are allowed to play their own life, even if limited. 

So, if we start with isotropic conditions with weak turbulence conditions $\chi \ll 1$, either there is a strict barrier to a parallel cascade due to the removal of the nonlinear parallel coupling as in RMHD, or (in the full MHD case), there is no barrier, and the scenario proposed here is allowed to develop the GM regiime.

In conclusion, weak turbulence appears to come in three flavors in MHD: (i) the isotropic IK phenomenology which holds in 2D MHD (ii) the strongly anisotropic version with no parallel interactions (iii) the mildly anisotropic GM regime studied here.
The first and third versions show $k^{-3/2}$ spectra. The last regime is to be found only in MHD. The second regime is probably to be found as well in MHD and RMHD.
We don't know yet what determines the transition between the second and third regimes in MHD. This issue is a work in progress.

\begin{acknowledgments}
We thank G. Belmont, Y. Dong and A. Verdini for several fruitful discussions.
\end{acknowledgments}

\bibliography{grappin}

\end{document}